# The Detection of Unconventional Quantum Oscillations in Insulating 2D Materials

Sanfeng Wu

Department of Physics, Princeton University, Princeton, New Jersey 08544, USA
* Email: sanfengw@princeton.edu

**Abstract -** In strongly correlated quantum materials, electrons behave in ways that often extend beyond the confines of conventional Fermi-liquid theory. Interesting results include the observation of low-temperature metallic behavior in systems that are highly resistive. Here we provide an overview of experiments in which insulators exhibit characteristics of a metal such as the Shubnikov–de Haas-like quantum oscillations, focusing on recent findings in the correlated insulating states of two-dimensional $WTe_2$. We discuss the status of current research, clarify the debates and challenges in interpreting the experiments, rule out extrinsic explanations and discuss promising future directions.

## I.    Introduction

The distinction between metallic and insulating electronic phases in crystalline materials is one of the most important achievements of the quantum theory of solids. In crystals, electrons obey the Pauli exclusion principle and occupy energy bands in accordance with Fermi-Dirac statistics. In metals, partial occupation of a band leads to the existence of a Fermi surface. In insulators, however, a well-defined energy gap separates fully occupied bands from unoccupied bands. Electronic phases can hence be classified by the absence (insulator), or presence (metal), of a finite Fermi surface with gapless electronic excitations. The defining characteristics of a metal include high electrical conductivity down to cryogenic temperatures, excellent screening of electric fields, and the Shubnikov–de Haas (SdH) and de Haas-van Alphen quantum oscillations (QOs) in magnetic fields ($B$). These features are not expected in insulators (table 1).

There are however strong motivations to search for unconventional insulators that host phenomena featuring characteristics typically belonging to a metal. For instance, theoretical approaches to the quantum spin liquid problem suggest how electrons in strongly correlated insulators may fractionalize into charge-neutral quasiparticles (spinons) which occupy states enclosed by a neutral Fermi surface[1,2]. This new type of insulators, if exist, should exhibit characteristics of metals, despite being highly resistive to an applied charge current. Experimental detection of a neutral Fermi surface inside a charge gap is however extremely challenging and so far, inconclusive[1,2]. Among many approaches, the detection of intrinsic SdH-like QOs, with periodicity in $1/B$, in an insulator holds promise because of its potential connection to not only the Fermi surface but also the emergent gauge fields that is believed to be essential in the fractionalized phase[3,4].

| Properties\Phases | Insulator | Metal |
|---|---|---|
| **Electrical conductivity** | vanishing | high |
| **Electric-field screening** | weak | strong |
| **Electronic excitations** | gapped | gapless |
| **Fermi surface** | no | yes |
| **QOs or QHEs** | no | yes |

**Table 1. Characteristic features contrasting insulators and metals in established theory.**

QOs in insulating materials have been previously discussed in a few systems, such as Kondo insulators $SmB_6$[5,6] and $YbB_{12}$[7], topological excitonic insulator $WTe_2$[8], and the quantum spin liquid candidate α-RuCl$_3$[9]. However, debates on the nature of the observed QOs are ongoing[10–21]. The roles of surface states[10] and metallic impurities[22,23] in generating QOs in Kondo insulators have been discussed extensively. The nature of the insulator phase and low energy excitations (fermionic, bosonic or anyonic) of α-RuCl$_3$ remains largely unsettled at this moment[18–21,24]. In the case of WTe$_2$, in which a series of its intriguing quantum phases have been revealed recently, the puzzle of Landau quantization is an outstanding unsolved problem. To advance the field, it is essential to firmly establish the existence of one example of a strongly insulating material that exhibits some of the quantum characteristics of a metal. In this perspective, we highlight the promise of WTe$_2$ and subsequently strongly correlated insulators in 2D materials and van der Waals (vdW) moiré stacks.

## II.  Selected Theoretical Possibilities of Intrinsic QOs in 2D Insulating Materials

When SdH-like QOs and sharp conductance peaks akin to Landau quantization were first reported in monolayer WTe$_2$ insulators[8], there was no prior theoretical prediction. It was a surprising observation. In subsequent years, a number of theoretical proposals have been applied to the results in WTe$_2$. We briefly summarize some of proposed scenarios. (1) Gap modulation in a topological excitonic insulator (EI) – In a class of small-gap insulators arising from hybridization of electron and hole bands, theories have shown that the gap size may develop oscillations in magnetic fields[12,13,25]. This scenario has been explored in the EI state of monolayer WTe$_2$, predicting QOs in the thermally activated electrical conductivity by Lee[17,26] (Fig. 1a). (2) Charge-neutral Fermi surfaces - An intriguing scenario is the emergence of charge-neutral Fermi surface in the insulator state (Fig. 1b). In this case, spin-charge separation of hole quasiparticles leads to fractionalized holons and spinons. The latter follows Fermi statistics and may couple to external magnetic fields via internal gauge fields[4], like that proposed in the gapless quantum spin liquid[1,27] and mixed-valence insulators[28,29]. Landau quantization of the spinon Fermi surface is proposed to give rises to QOs in measurable quantities of the insulator[4,29–31]. (3) Recent experiments on small-angle twisted bilayer WTe$_2$ (tWTe$_2$)[32–34] have observed a 2D anisotropic phase akin to a Luttinger liquid (LL), characterized by an exceptionally large in-plane transport anisotropy, a power-law scaled conductance in the hard transport direction and a distinct nonlinear differential resistance featuring a zero-bias dip in the easy direction. In this non-Fermi liquid 2D phase, transport in the hard direction follows a power law and vanishes at low temperatures ($T$), i.e., being insulator-like, although the underlying phase is gapless. Magneto-transport of this 2D anisotropic LL is highly interesting and remains to be explored. Theoretical explorations, in the context of coupled wire constructions, suggest possibilities of QOs and novel quantum Hall effects (QHEs) (Fig. 1c)[35–38]. Interestingly, spin-charge separation is naturally expected in a LL. Hence how it manifests in such a 2D anisotropic setting deserves careful theoretical and experimental studies, the exploration of which might be of relevance to scenario (2).

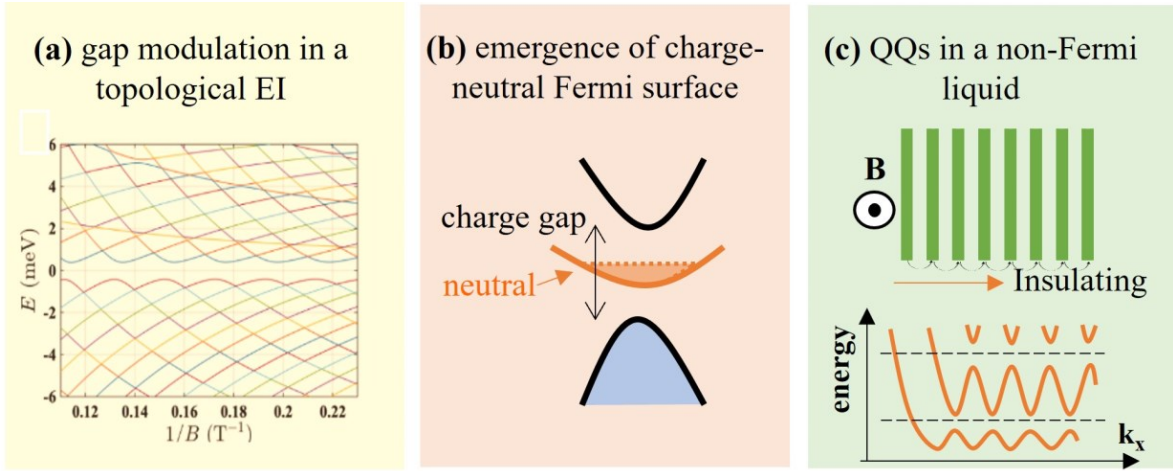

**Fig. 1. Selected intrinsic mechanisms proposed for observing QOs in 2D insulators. (a)** QOs of the insulating gap**.** The diagram is reproduced from ref. 26, adapted by permission from American Physical Society [26] **(b)** Emergent charge neutral fermions and a Fermi surface inside a charge gap. **(c)** An illustration of QOs in a non-Fermi liquid (2D anisotropic LL).

### III. Transport Observation of Landau Quantization in Insulating WTe$_2$ Monolayers

Experimentally, few layer WTe$_2$ is a semimetal but monolayer WTe$_2$ develops a strongly insulating state at low $T$ both around the charge neutrality point (CNP) intrinsically and in the wide hole-doped range in the presence of disorders[8,39–41]. This is in sharp contrast to the electron-doped regime, where a metal-insulator transition is seen typically near a gate-induced electron doping of ~ $10^{12}$ cm$^{-2}$. With electron doping, the monolayer also becomes a superconductor at sub-kelvin temperatures[42–44]. However, the hole-doped side is different as the monolayer remains an insulator even with significant amount of doping, presumably due to localization effects. Striking QOs, as well as sharp conductance peaks mimicking the conventional Landau quantization features in 2D electron gas, are surprisingly observed both in the metallic and in the insulating phases of the monolayer, with many oscillating periods onset at low $B$ of less than 1 T (Fig. 2a)[8]. These observations are highly reproducible in many devices. In relatively disordered samples, QOs in the insulating regime remain strong while they are barely visible in the metallic electron side. In cleaner samples, QOs are visible in both electron- (metal) and hole-doped (localized insulator) regimes, as well as near CNP (EI), as seen in Figs. 2b & c. More details of the experimental observations can be found in the reference[8].

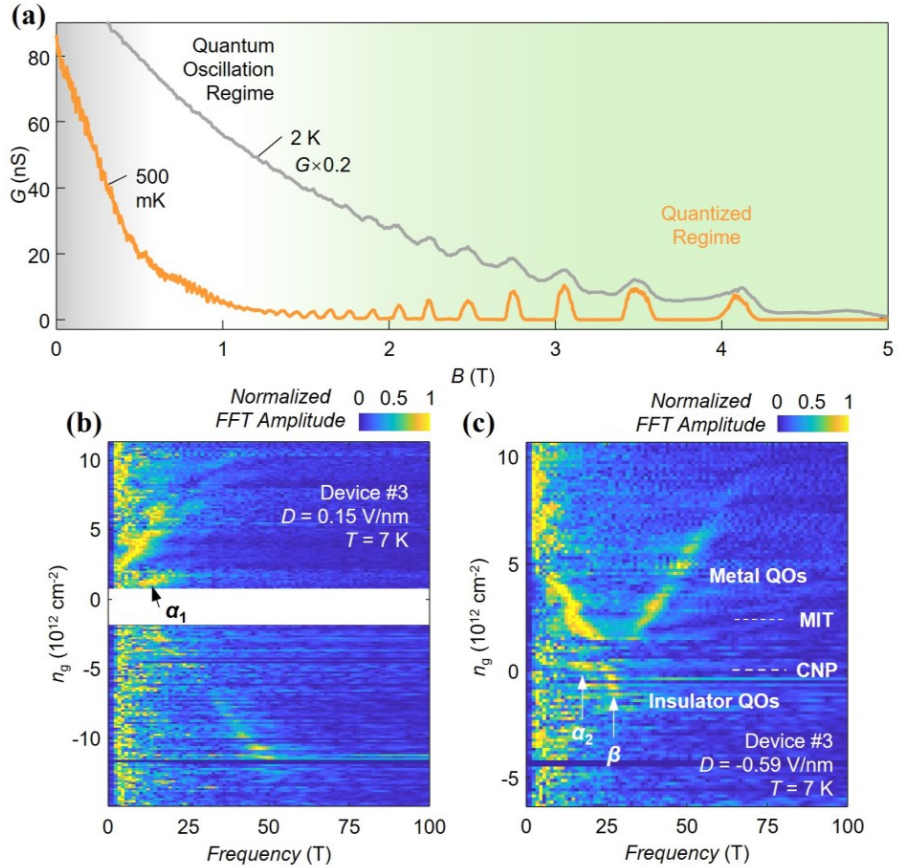

Fig. 2. Experimental observation of Landau quantization in the transport of WTe$_2$ monolayer insulator. **(a)** Magnetoresistance curves at 2 K and 500 mK taken in the insulating regime of a device. **(b)** Observation of multiple QO branches in a high-quality device, in three different regimes of electron-doped metal, hole-doped insulator and around CNP (insulator). Data replotted from Nature 2021 [8].

In our original report[8], different scenarios for interpreting the highly unexpected experimental results have been scrutinized. These include the most intriguing case where the presence of highly mobile fermionic quasiparticles in the insulator causes the QOs (Fig. 1b). Another scenario is the *B*-induced gap modulation that was initially proposed in the context of Kondo insulators[12,13]. A similar scenario was later applied by Lee to $WTe_2$-like excitonic insulator phase[17,26], as mentioned above (Fig. 1a). The difficulties in this picture lie in the explanation of the low onset field (suggestive of highly mobile quasiparticles), the need of the exponential form of the thermally activated conductivity (which is typically deviated at low *T* in the real devices) and the widely present QOs away from CNP on the hole-doped side (presumably a localized insulator in the presence of disorders). Another alternative explanation is to attribute the observed QOs to the nearby graphite gate used in the devices. This scenario has been carefully discussed in the method section of our original experimental report[8], and was later expanded by Mak et al[16], in which they proposed a "unified mechanism" that attributes all observed QOs in insulating 2D materials in the presence of a graphite gate to the graphite. Concerns regarding this work have been made previously in a comment by us[45]. In the next section, we highlight the issues associated with applying the graphite-gate mechanism in the $WTe_2$ case.

### IV. The Proposal of Graphite-Gate Scenario for $WTe_2$ QOs

We start by briefly clarifying the debate regarding the graphite-gate scenario. First, there is no disagreement that the potential impact of graphite QOs in graphite-gated devices should be carefully examined, as indeed done in the original report of $WTe_2$ monolayer QOs[8]. At the core of the debate is whether the graphite-gate scenario proposed in Mak et al[16] is truly consistent with the essential details of the experimental facts or not. Our conclusion is that the proposal presented in Mak et al[16] is inapplicable to the findings in WTe$_2$.

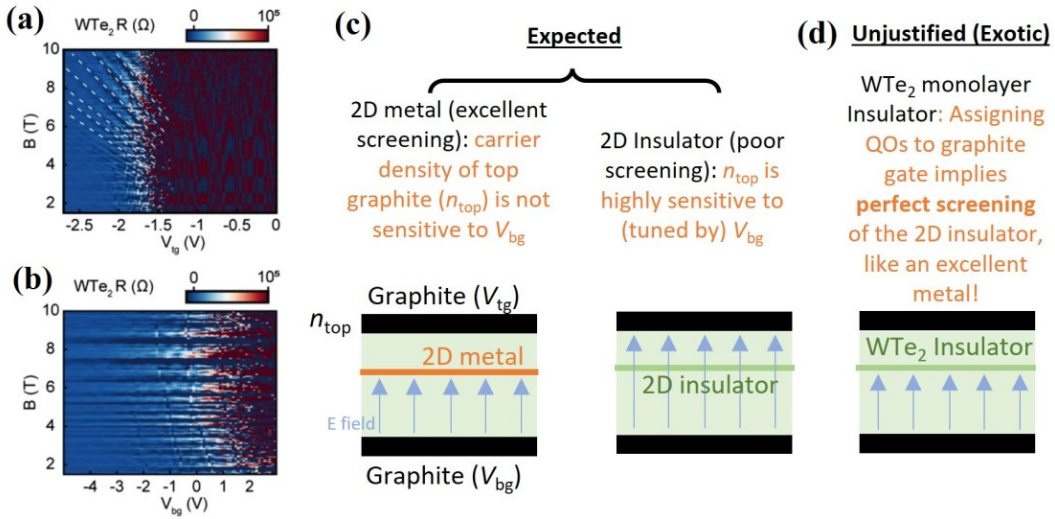

**Fig. 3. The unjustified and exotic proposal of graphite-gate scenario. (a & b)** QOs of monolayer $WTe_2$ insulator reported in Mak et al [16]. Adapted by permission from American Physical Society [16]. **(c)** Trivial expectation of the carrier density in the top gate. **(d)** An unjustified assumption in the graphite-gate scenario of Mak et al: the monolayer $WTe_2$ insulator screens the electric field perfectly like an excellent metal.

Here we highlight one highly nontrivial and unjustified aspect of Mak et al[16] proposal, considering their own data (Figs. 3a & b). Note that unlike our original report[8], where multiple QO branches were observed in three regimes (electron side, hole side and around CNP), Mak et al[16] reported only one branch of QOs on the hole-doped side extended to near CNP. As mentioned above, the wide hole-doping range of monolayer WTe$_2$ exhibits insulating behaviors. In Mak et al, the aspect that the observed QOs only depend on the top gate voltage but not the bottom gate voltage was regarded as a key evidence for assigning the QOs to the top graphite gate. This argument, however, requires a highly nontrivial assumption that the highly insulating WTe$_2$ monolayer (both on the hole side and near CNP) screens the electric field perfectly like an excellent metal! The expectation that the carrier density of top graphite becomes independent of the bottom gate voltages is only trivial when the 2D layer is a good, high density metal. In the case of an insulator, the carrier density of top graphite should be, on the contrary, highly sensitive to the bottom gate (the two gates form a capacitor). We illustrate this important aspect in Figs. 3c & d. As seen from their data (Fig. 3b), this perfect screening of electric fields needs to be true even near CNP, where the monolayer resistance is so large that it can't even be measured reliably, an assumption left unjustified at all in Mak et al[16]. We note that even in this seemingly trivial graphite-gate scenario, the WTe$_2$ monolayer insulator still exhibits highly exotic properties, behaving like a metal that violates conventional wisdom.

There are other unjustified aspects of the Mak et al proposal[16], as pointed out in our previous comment[45]. For instance, Mak et al argued that a rapid change of resistance $R$ under varying carrier density $n$, i.e., a large $dR/dn$, is necessary in their proposed mechanism. However, the strongest QOs observed WTe$_2$ occurred at the most insulating CNP and the hole doped side, where $dR/dn$ was mostly suppressed (as also seen in the data of Mak et al). In fact, $dR/dn$ of WTe$_2$ monolayer peaks at slight electron doping, yet QOs there are often not seen or very weak. Mak et al[16] also highlighted that a thin top hBN (~ 5 nm) plays an important role in the observation, but we don't find it critical. In addition, our original report revealed much more information, including more QO branches in different regimes, whose behaviors are inconsistent with the proposed graphite-gate mechanism.

## V. Efforts towards High-Quality Devices without Using Graphite Gates

In this section, we report on our ongoing efforts to detect QOs of 2D WTe$_2$ systems without using graphite gates. Unlike graphene or a few other semiconducting 2D chalcogenides (e.g., MoS$_2$), WTe$_2$ monolayers are sensitive to environments and can easily degrade if care is not well taken during the fabrication process. Significant challenges are thus present in preparing high quality devices necessary for observing QOs. The challenges are also reflected by the fact that QOs of monolayer WTe$_2$ were not reported after several years since its observation of the robust topological edge modes. The optimized fabrication procedure and the employment of double graphite gate structure in the reference[8] has improved the device quality significantly, which allowed for the first observation of QOs in monolayer WTe$_2$. In the following few years, we have spent substantial efforts in improving the quality of WTe$_2$ devices using only elemental metal, instead of graphite, as gates. The motivation is that QOs expected in elemental metal are of much higher frequencies well separated from the WTe$_2$ QOs, and hence such a device will automatically exclude the possibilities of assigning the low frequency QOs to the nearby gate. The roughness of deposited metal is, however, much higher compared to the atomically flat graphite, and introduces substantial sources of disorders. The constraint that monolayer WTe$_2$ can only stay in glovebox during the fabrication, even when it is encapsulated in a vdW stack, provides additional challenges for fabricating metal-gated devices. Nevertheless, we have managed to produce metal-gated monolayer devices of reasonably good qualities that reproduced the observations of e.g., topological excitonic insulator state and gate-induced superconductivity. The latest details of how we achieved the fabrication of double metal (Pd) gated 2D WTe$_2$ devices can be found in our recent paper[33], in which we successfully developed such devices and observed the 2D anisotropic Luttinger liquid phase in twisted bilayer WTe$_2$ down to millikelvin temperatures.

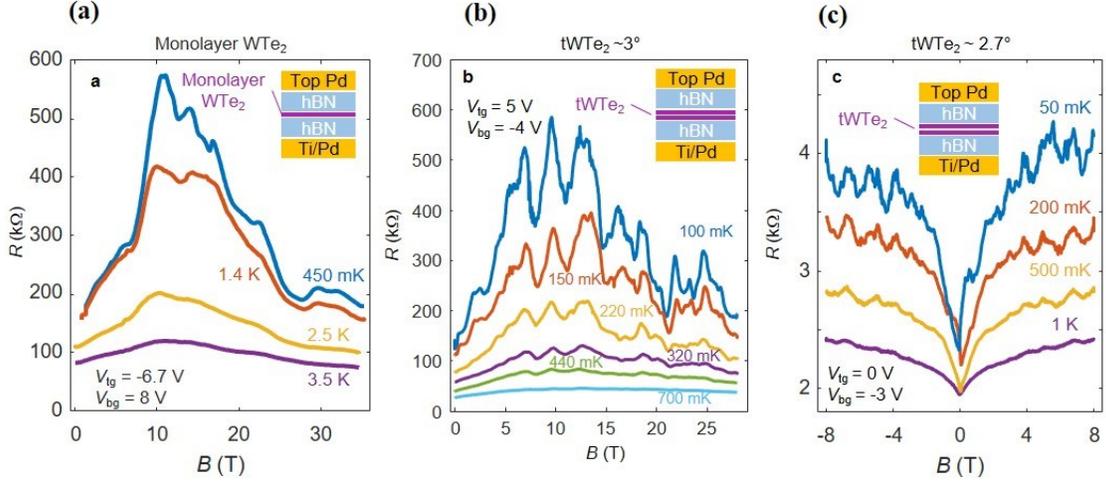

**Fig. 4. QOs observed in metal (Pd)-gated WTe$_2$ devices. (a)** Magnetoresistance of a monolayer WTe$_2$ device taken at different temperatures. **(b & c)** Magnetoresistance of twisted bilayer WTe$_2$ devices (**b**, with an interlayer twist angle of ~ 3°, and **c**, 2.7°) taken at different temperatures.

Progress has also been made in observing unconventional QOs in 2D WTe$_2$, which requires even higher device quality compared to that for observing other properties. Particularly, we have seen in these metal-gated devices sdH-like oscillations, with periodicity in $1/B$, in the magnetoresistance of monolayer WTe$_2$ (Fig. 4a) and small angle twisted bilayer WTe$_2$ (Figs. 4b & c) at selected gate voltages, where the resistance increases with decreasing temperature (i.e., exhibiting insulating behaviors). These QOs in the apparently insulating state are unexpected in conventional Fermi liquid picture. We also emphasize a major remaining obstacle in comparing these data with graphite-gated devices, which is the gate-voltage dependence of these oscillations. In the metal-gated devices, we often saw extra bumps and distortions in the magnetoresistance curves that can be attributed to disorders, which appear to be coupled to the gate voltages. As a result, a change in the gate voltage will introduce extra features that distort the QO features, preventing us from extracting reliable gate dependent behaviors of QOs. A related challenge is that, with lower quality compared to graphite-gated device, one needs higher $B$ to resolve QOs. Indeed, the data shown in Figs. 4a & b were taken in National High Magnetic Field Laboratories (NHMFL). The time constraints there, compounded by difficulties working with air-sensitive materials, prevent us from systematically measuring and understanding the QOs in these

## VI. New Experimental Insights: Sign-Change Thermoelectric QOs

Removing the graphite gate from the device is not the only route to unambiguously rule out the graphite-gate interpretation of the QOs. Here we discuss a new set of experiments on monolayer WTe$_2$, which clarify the situation considerably. These recent results also move the field of QOs in insulators into a new regime.

QOs can be detected in various properties of the materials, including e.g., resistivity, magnetization, and thermoelectricity. Among those, thermoelectricity, such as the Seebeck effect, is special since the thermopower is sensitive to the energy derivative of the density of state near the chemical potential, rather than the carrier density itself. Particularly, the thermopower can be of different sign, reflecting the type of carriers diffusing from the hot to the cold end being electron-like or hole-like. For instance, in magnetic fields the Landau quantization of 2D electron gas would produce alternating sign-change thermopower, signifying an energy level (i.e., the Landau level) passing through the chemical potential (Fig. 5a). Indeed, this sign-change thermoelectric QOs due to Landau quantization has been observed in graphene[46,47].

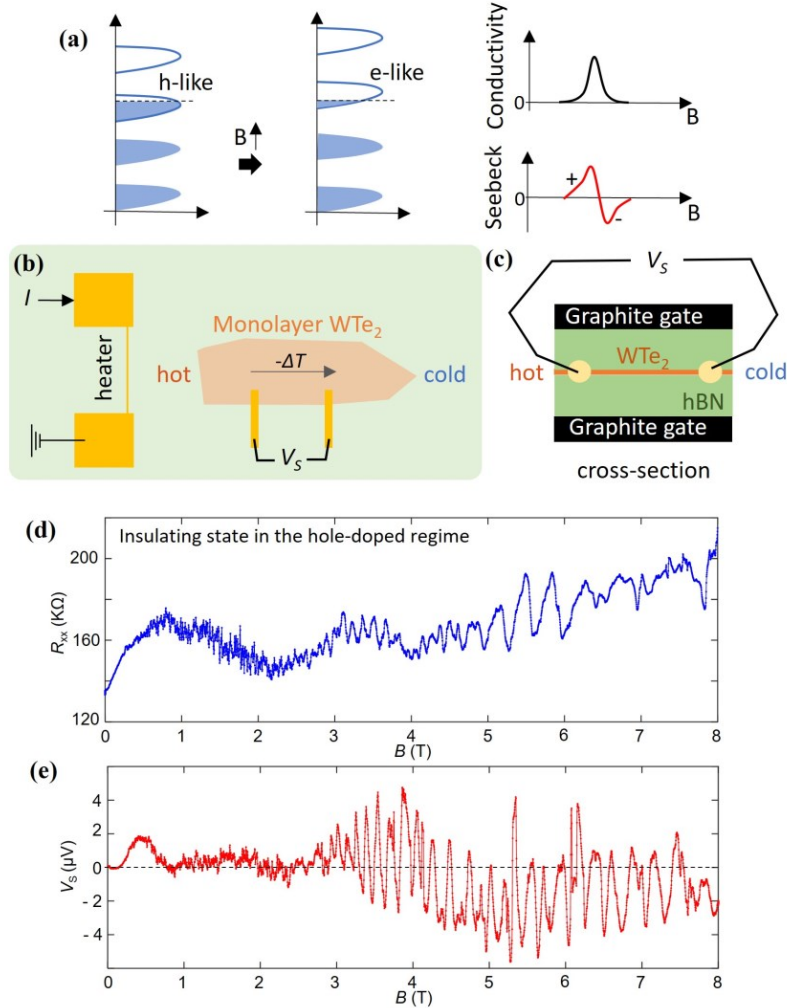

**Fig. 5. Sign-change thermoelectric QOs observed in monolayer WTe$_2$.** **(a)** Illustration of effects of Landau quantization on conductance and the Seebeck effect, respectively, of a conventional 2D electron gas. **(b)** Cartoon illustration of the measurement scheme for Seebeck effect in monolayer WTe$_2$. **(c)** Cross-section view of the device design. **(d)** Magnetoresistance of monolayer WTe$_2$ taken in the hole doped insulating regime, exhibiting QOs. **(e)** The Seebeck voltage ($V_S$) recorded in the thermoelectricity measurement (i.e., there is no external current applied to WTe$_2$; only a temperature gradient is applied) on the same device and at the same applied gate voltages. The Seebeck voltage exhibits QOs consistent with the transport measurement and, importantly, display clear sign-changes. Details of the experiments are included in a separate manuscript [48].

devices. Nevertheless, we believe there is still room to further improve the quality of metal-gated WTe$_2$ device, an ongoing effort for its direct comparison with graphite-gated devices.

We performed thermoelectricity measurements in monolayer $WTe_2$. Cartoon illustration of the detection scheme is shown in Figs. 5b & c. Details of the device structure and fabrication were described in our recent work in which we measured the Nernst effect in the superconducting state of monolayer $WTe_2$[44]. Here we focus on the Seebeck effect of the monolayer in the insulating hole-doped regime. Fig. 5d plots a typical magnetoresistance and QOs in the insulating regime at a selected gate voltage, consistent with our previous transport report. Fig. 5e displays the Seebeck voltage ($V_S$) measured between two open contacts placed in the longitudinal direction when a temperature gradient ($\Delta T$) is applied to the device. Dramatic QOs, consistent with the resistance QOs, are seen in the thermoelectric voltage and remarkably, they clearly display sign changes under varying magnetic fields! Similarly, if one shorts the two contacts and monitors the electrical current generated by the $\Delta T$ a sign-change alternating thermoelectric current flowing in the $WTe_2$ is also observed. Such behaviors are consistently observed over a wide hole-doped insulating regime extended to near CNP. More details of these intriguing observations are discussed in a separate manuscript[48].

The sign-change thermoelectric QOs induced by magnetic fields is inconsistent with any scenarios that attribute the QOs to small changes of carrier density, such as the graphite-gate scenario. A small carrier density modulation of $WTe_2$ induced by the graphite QOs, as proposed in Mak et al[16], cannot generate a sign change of the thermoelectric response in the $WTe_2$ layer. As noted before, the sign-change in the thermopower implies a qualitative change of the carrier type in the $WTe_2$ monolayer from being electron-like to hole-like! We are not aware of any feasible mechanism under which this can be caused by the graphite gate. The sign-change thermoelectric QOs are however consistent with the presence of Landau-level-like energy structure in the insulating state of monolayer $WTe_2$, reinforcing the scenario of neutral fermions[8].

**Summary and Outlook**

The unexpected observation of QOs in monolayer $WTe_2$ remains an interesting and challenging puzzle. We hope the community continues to suggest or consolidate possible mechanisms and propose experiments that can test them. We believe that the solution to the intriguing Landau quantization problem of monolayer $WTe_2$, a material that has produced so many surprises on multiple frontiers, likely holds a key to new discoveries in quantum materials.

Behind the puzzle are the experimental challenges in detecting quantum properties of insulators and the promise of addressing them. Indeed, motivated by the observation of $WTe_2$ QOs, we are now in the process of developing and optimizing new experimental tools for probing charge neutral excitations in 2D insulators. The monolayer thermoelectricity described above is one example. The theory and experiments of thermoelectricity in spin-charge separated 2D systems deserve careful studies. Such theoretical developments will be important in further understanding the experiments. Another direction to explore is to look for signatures of neutral fermionic excitations using optical spectroscopy, which can in principle detect charge-neutral modes below the charge gap of an insulator. The recent advancement of a new instrumental platform enabling far-infrared optical spectroscopy of 2D materials devices at millikelvin temperatures[49] can be helpful. In particular, the detection of sub-gap optical absorption[50], including e.g., cyclotron resonance[51], could be important in determining the nature of low-energy excitations. Other techniques, such as nonlocal spin-transport, may also be useful in revealing properties of spinons if they are present. It is important to note that while 2D $WTe_2$ provides a testbed for experimentally examining novel ideas (e.g., charge-neutral Fermi surfaces and spin-charge separation) and for searching for unimagined phenomena, the experimental techniques developed for this endeavor can and will also be applied to studying a variety of other correlated 2D crystals and moiré materials. We believe such efforts will be fruitful in leading us to uncharted territories in strongly correlated quantum matter.

**Acknowledgements**

We acknowledge discussions with N. P. Ong and A. Yazdani and experiments performed by Pengjie Wang, Guo Yu, Tiancheng Song and Yue Tang. This work is supported by NSF through a CAREER award (DMR-1942942) and the Materials Research Science and Engineering Center (MRSEC) program of the National Science Foundation (DMR-2011750), ONR through a Young Investigator Award (N00014-21-1-2804), AFOSR Young Investigator Award (FA9550-23-1-0140), the Gordon and Betty Moore Foundation through Grants GBMF11946 and the Sloan Foundation. A portion of this work was performed at the National High Magnetic Field Laboratory, which is supported by National Science Foundation Cooperative Agreement No. DMR-2128556 and the State of Florida.

**Competing interests**

The author declares that he has no competing interests.